\documentclass{sf2a-conf2013}
\usepackage{graphicx}
\usepackage{hyperref}
\usepackage[]{natbib}  
\usepackage{epstopdf}

\def\BibTeX{{\rm B\kern-.05em{\sc i\kern-.025em b}\kern-.08em
    T\kern-.1667em\lower.7ex\hbox{E}\kern-.125emX}}
\bibpunct{(}{)}{;}{a}{}{,}  

\begin{document}

\TitreGlobal{SF2A 2013}


\title{High Precision Astrometry in Asteroid Mitigation - the NEOShield perspective}

\runningtitle{NEOShield}

\author{S. Eggl}\address{IMCCE - Observatoire de Paris, 77 Avenue Denfert-Rochereau, 75014 Paris, France}

\author{A. Ivantsov$^1$}

\author{D. Hestroffer$^1$}

\author{D. Perna}\address{LESIA - Observatoire de Paris, 5, place Jules Janssen
92195 MEUDON Cedex, France}

\author{D. Bancelin$^{1,}$}\address{Institute for Astrophysics, University of Vienna, T\"urkenschanzstr. 17., 1180 Vienna, Austria}

\author{W. Thuillot$^1$}




\setcounter{page}{170}


\maketitle


\begin{abstract}
Among the currently known Near Earth Objects (NEOs), roughly 1400 are classified as being 
potentially hazardous asteroids. The recent Chelyabinsk event has shown that these objects can pose a real threat to mankind. 
We illustrate that high precision asteroid astrometry plays a vital role in determining potential impact risks, 
selecting targets for deflection demonstration missions and evaluating mitigation mission success.  
After a brief introduction to the NEOShield project, an international effort initiated by the European Commission to investigate aspects of NEO mitigation in a comprehensive fashion, 
we discuss current astrometric performances, requirements and possible issues with NEO risk assessment and deflection demonstration missions.    
\end{abstract}



\section{Introduction}
Catastrophic collisions of our Earth with asteroids larger than 10km are fortunately rare, yet impact frequencies for objects sized
between 15m and 500m are non negligible, as the recent event in Chelyabinsk has shown \citep{harris-2010}. Given the fact that roughly 
1400 among the currently known Near Earth
Objects (NEOs) are potentially hazardous asteroids (PHA), a comprehensive understanding of the orbital and physical properties of these
celestial bodies is desirable in order to allow for a reliable threat assessment.
The NEOShield project \citep{harris-et-al-2013} constitutes an international effort under European leadership to enhance our understanding
of threat mitigation options regarding potential asteroid impacts. In the framework of this project different mitigation strategies, such as
blast deflection, kinetic impactors or gravity tractors are not only studied individually, but they are put in context of realistic
threat scenarios. Past missions to Comets and asteroids such as Rosetta (ESA), Hayabusa (JAXA) and NASA's DAWN and Deep Impact missions have shown 
that we are capable of performing rendez-vous as well as high velocity collisions with a minor planet. Many open questions remain, however, 
regarding the details of asteroid deflection, such as the momentum enhancement due to ejecta, optimum targeting and impacting strategies.  
The development of a mitigation test mission will, thus, be a focus of the NEOShield project.

Asteroid deflection might seem to be centered on engineering issues. Yet, the importance of high precision astrometry should not be underestimated in this context. 
In the following sections we will discuss multiple topics that require
high quality astrometric data in order to provide vital input to asteroid threat assessment and mitigation evaluation.
  
\section{NEO Threat Assessment}
\label{eggl:sec2}
Given the fact that our knowledge of the exact orbit of a NEO at any given
time is limited by observational as well as modeling uncertainties, threat
assessment has to be conducted on a probabilistic basis. 
Hereby, the standard approach is to perform an initial orbit determination using a small subset of the available NEO
observations. A differential correction method is applied to improve the NEO's
orbit using weighted residuals of all available observation data  \citep[e.g.][]{milani-gronchi-2010}. 
This yields a covariance matrix that contains the observation uncertainties of the nominal trajectory. The covariance matrix of an
asteroid's orbital elements at a given epoch is then used to generate initial state
vectors of virtual asteroids (VA) or clones following a line of variation (LOV) sampling
approach \citep{milani-et-al-2005, milani-et-al-2002}. The equations of motion of the
nominal asteroid and its clones are solved numerically and their trajectories are
monitored for close encounters with the Earth. Should a close encounter occur, then
the minimum encounter distance is calculated, and the target plane (b-plane)
coordinates of the respective clone are stored together with the minimum
encounter distance. Potential virtual impactors (VI) are flagged and the impact probability
can be calculated following e.g. \citep{milani-et-al-2005, valsecchi-et-al-2003}.

The fact that the uncertainty in an NEO's orbit plays an essential role in the impact prediction machinery makes it necessary to  
have observational data of sufficient quantity and quality to perform a reliable threat assessment.
In other words, the better constrained the initial orbit uncertainties, the more accurate the impact probability estimates.
This, of course, provokes the following questions: How well constrained should NEO orbits be to allow for a reliable impact analysis? How many observations
are necessary to achieve such constraints?

Global answers to these questions
are difficult to obtain, since the achievable improvement depends on many parameters, such as the quantity and quality of previous observations, weighting of residuals
the length of the achieved data arc, and so on \citep{carpino-et-al-2003,baer-et-al-2011}.
Radar observations usually are very valuable in this respect, since range and range-rate measurements are not readily accessible via optical astrometry, unless 
triangulation can be applied \citep{eggl-2011}.
Due to the relatively short range and high cost of radar observations, orbit determination and improvement of asteroids is still performed primarily via optical astrometry. 
The quality of the latter, thus, plays an essential role in determining whether or not an asteroid poses a potential threat to our planet.

\begin{figure}[ht!]
 \centering
 \includegraphics[width=0.7\textwidth,clip]{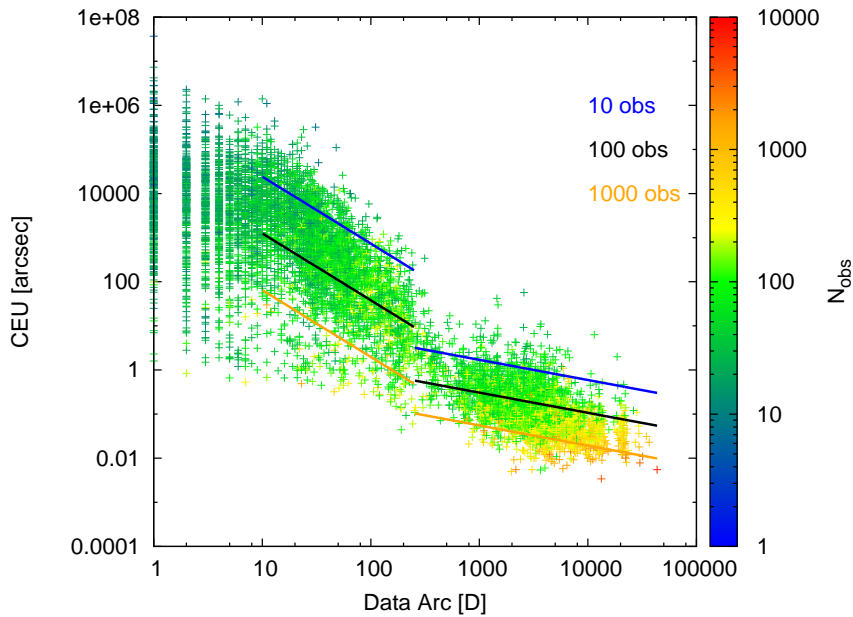}      
  \caption{Current Ephemeris Uncertainty (CEU) of NEOs as a function of the observed arc-length and the number of observations. While the quality of initial orbit determination for very short arcs ($< 10$ days) can only be evaluated
  on a case to case basis, trends in single opposition NEOs ($10 < arc < 200$ days) and multi-opposition NEOs ($arc > 360$ days) can be crudely estimated by log-linear least squares fits.}
  \label{eggl:fig1}
\end{figure}

While precise numbers are hard to come by, a rough correlation between the expected orbit quality and the available quantity of observation data can be inferred from   
plotting the Current Ephemeris Uncertainty (CEU), which is basically the positioning uncertainty in an observer's sky plane 
\citep{bowell-2009}, against the number and arc-length of observations \citep{desmars-et-al-2013}, see Fig.~\ref{eggl:fig1}.\footnote{The values shown in Fig.~\ref{eggl:fig1} have been generated using the currently discontinued astorb database as of 10/2012.}.

\begin{figure}[ht!]
 \centering
 \includegraphics[width=0.7\textwidth,clip]{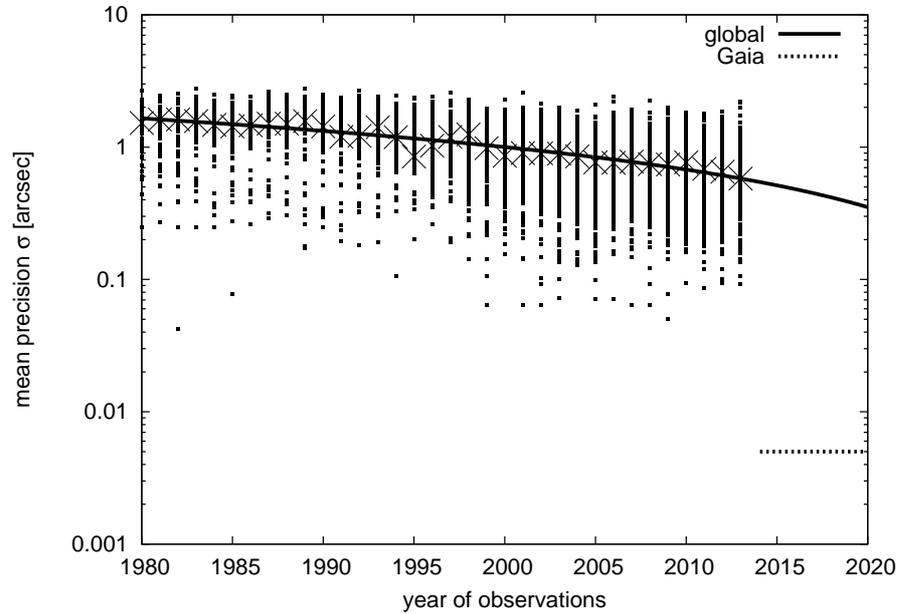}      
  \caption{The yearly astrometric mean precision ($\sigma=\sqrt{(\sigma_\alpha \cos{\delta})^2+(\sigma_\delta)^2}$) of each observatory with a Minor Planet Center designation and at least 10 submitted observations per year is plotted against time (dots).
  The crosses indicate global yearly averages, i.e. the weighted precision of all observations considered. The full line indicates the least squares trend of the global yearly averages, 
  i.e. the mean quality of astrometric measurements. The dashed line represents pessimistic estimates for the performance of ESA's Gaia mission \citep{tanga-2012}.}
  \label{eggl:fig2}
\end{figure}

One can see that estimating the accuracy of initial orbit determination is difficult,
especially, when only short observational arcs ($ < 10$ days) are available. The transition region from single to multi-opposition appears as a gap
around 200 days. In order to get a rough idea on the CEU improvement with arc-length and number of observations, we can construct two log-linear least squares fits
for single and multi-opposition domains
\begin{equation}
\log_{10}CEU=a_1+a_2\; \log_{10}\;arc+a_3\; \log_{10}\;N_{obs}. \label{eggl:fit1}
\end{equation}
The corresponding coefficients are given in Tab.~\ref{eggl:tab1}.
\begin{table}
\begin{center}
\begin{tabular}{cccc}
single opposition: & $a_1=7.08\pm0.09$ & $a_2=-1.83\pm0.04$ & $a_3=-0.93\pm0.05$\\
multi-opposition: & $a_1=2.34\pm0.07$ & $a_2=-0.46\pm0.02$ & $a_3=-0.74\pm0.02$.\\
\end{tabular}
\end{center}
\caption{Coefficients for the log-linear least squares fits (\ref{eggl:fit1}) shown in Fig.~\ref{eggl:fig1}.\label{eggl:tab1}}
\end{table}
It is also evident from Fig.~\ref{eggl:fig1} that CEU values $\ll 1''$ are hard to achieve unless
the NEO has been observed on a regular basis as well as over time spans that cover multiple oppositions. The reason for this apparent limit becomes clear when we 
consider the average quality of astrometric observations over the past years (Fig.~\ref{eggl:fig2}).

Here, the yearly astrometric mean precision of each observatory with a Minor Planet Center designation is plotted as a function of time.
However, observatories with less than 10 submitted observations per year have been omitted in order to guarantee the validity of the residual statistics.
The astrometric mean precision ($\sigma$) is defined as $\sigma=\sqrt{(\sigma_\alpha \cos{\delta})^2+(\sigma_\delta)^2}$, where $\sigma_{\alpha,\delta}$ are the yearly mean precisions of all astrometric measurements
of asteroids performed by the respective observatory.\footnote{\url{http://www.minorplanetcenter.net/iau/special/residuals.txt}}.
A weighted average using the respective number of considered residuals per observatory yields the global astrometric means denoted by the crosses in Fig.\ref{eggl:fig1}. 
The linear least squares fit of the latter data can then be used as a crude indicator of the average astrometric performance to be expected given a specific year:
$\bar{\sigma}=b_1+b_2\; t$ with $b_1=65.932\pm 3.619$, $b_2=-0.032\pm0.002$ and $[t]=[yrs]$.  
It is evident from Fig.~\ref{eggl:fig2} that the average astrometric accuracy during the period where most of the NEOs have been discovered ranges between $0.1-3''$ with a sightly improving trend.
This mirrors the fact that the newly discovered NEOs tend to have CEU values of a similar magnitude (see Fig.\ref{eggl:fig1}).
While high precision astrometry measurements ($\sigma_{\alpha,\delta} <0.1''$) have been available on a regular basis for about 10 years, their impact on the global NEO astrometric performance is still small.
In fact, 98\% of the observations in 2013 have a precision in the range of $0.13-2.01''$.
We also find that current astrometric accuracy is within the same order of magnitude. 
ESA's Gaia mission is bound to change that. By generating a new celestial reference frame Gaia will solve many of the current problems related to
biases in astrometric catalogs \citep{chesley-et-al-2010,mignard-klioner-2012}, thus improving the accuracy of ground-based observations.  
It will also pave the way for unprecedented astrometrical measurement precision with many possible applications to
Solar System and asteroid science \citep{bancelin-et-al-2010,beauvalet-el-al-2012,delbo-et-al-2012,hestroffer-et-al-2010,ivantsov-et-al-2013,tsiganis-et-al-2012}. 
Pessimistic estimates on Gaia's 
astrometric precision are represented by the dashed lines in Fig.~\ref{eggl:fig2}.\footnote{The astrometric performance was estimated for $V\simeq20$ objects with an additional penalty factor of 5 due to more rapid proper motion of NEOs \citep{tanga-2012}.}  
It is easily seen that even under worst case assumptions Gaia will produce high precision astrometry that is two orders of magnitude beyond the global trend.
Even-though this sounds promising, the actual performance of the astrometric orbit improvement has to be studied on a case-by-case basis. \citet{bancelin-et-al-2012} have investigated the likely impact of direct observations by Gaia on the 
orbit of 99942 Apophis. Their findings suggest a possible decrease of the asteroid's orbit uncertainty by two orders of magnitude, not accounting for the additional benefits of the updated catalog.
Assuming such a drastic improvement of orbit uncertainties is indeed possible, 
 Gaia will contribute significantly to NEO threat assessment.

\section{Follow Up}
It is well known that NEOs populate a very active dynamical region in the Solar System. Close encounters with the terrestrial planets, non gravitational forces and
interaction with resonances cause their orbits to become chaotic on rather short time-scales with Lyapunov times of no more than a hundred years \citep{michel-1997,michel-et-al-2005,wiegert-et-al-1998}.  
A continuous update of the orbits of already discovered asteroids becomes, thus, necessary, if one seeks to avoid sudden 'disappearances'.
Fortunately, current computational facilities allow asteroids with well constrained orbits to be propagated accurately and thus remain retrievable over decades. 
The dynamical model that is used to predict future positions, however, has to be complete in the sense that all non-negligible forces are considered.
This is especially tricky with respect to unexpected non-gravitational forces \citep{jewitt-2012} or close encounters with main-belt asteroids.
The latter ones are often neglected in ephemeris calculations either for computational reasons, or because the masses involved are not well known.
Yet, a preliminary investigation by the authors shows that the Mars-crossing NEO 184266, for instance, will have a close encounter
with the main-belt asteroid 52 Europa in 2019. This close approach causes an observable deflection of roughly $0.06''$. If such a deflection is not accounted for, 
discrepancies between the predicted and the actual position of the NEO can grow very quickly. Thus, 
frequent observational updates using high precision astrometry can serve not only as a guarantee that known asteroids remain retrievable. 
Frequent measurements can also serve as a test for the quality of the applied dynamical model. Since Gaia will scan the whole sky several times during its 5 year mission,
high precision astrometric follow up of once observed objects is basically guaranteed. Given the 
relatively short life-time of the Gaia mission, however, one should start to consider a long term follow up program to continue high precision astrometry after the 
era of Gaia has come to an end.

\section{Deflection Validation}
So far we have discussed initial orbit determination and orbit correction as they constitute the main contributions of astrometry to NEO threat assessment. 
Yet, high precision astrometry can play another important role, namely in validating an asteroid's deflection.   
Sending a deflection mission to a potentially impacting asteroid could one day become a necessity to avert local or even global disaster. 
Therefore, the NEOShield consortium aims not only to produce a global response to such threat scenarios. The development of a deflection demonstration mission also constitutes
a substantial pillar of the project.  
Given the political issues involved in launching nuclear missiles as well as the long duration of continuous slow push methods such as the gravity tractor, kinetic impactors seem to be the most economic
choice for a deflection demonstration. A straight forward approach has been suggested in the Don-Quijote study \citep{harris-et-al-2006}. A space-craft (S/C) is launched and set on a collision course with a NEO. During the impact, the momentum of the S/C is transferred to the asteroid 
following the slightly adapted momentum conservation equation (\ref{eggl:eq1})
\begin{equation}
\Delta V=\beta \frac{m}{M} \Delta v, \label{eggl:eq1}
\end{equation}
where $\Delta V$ and $\Delta v$ are the changes in the heliocentric velocities of the asteroid and the S/C, $M$ and $m$ are the asteroid's and the S/C's masses, respectively, and  
$\beta$ denotes the momentum enhancement factor that is due to ejecta of surface material \citep{holsapple-housen-2012}.
Since it is rather expensive to launch a massive S/C with a substantial amount of fuel, test
deflections in the order of $\Delta V\simeq cm/s$ are being discussed. In order to being able to see the effect of such a minute push,
the asteroid's orbit has to be very well characterized before and after the deflection event.
In order to give a rough quantitative estimate on the necessary precision, we will assume that all of the momentum was used to change the NEO's semimajor axis ($a$).
Then we have \citep{vasile-colombo-2008}:
\begin{equation}
\Delta a =\frac{2 a^2}{\kappa} V \Delta V.
\end{equation}
Here, $\Delta a$ corresponds to the change in semimajor axis and $\kappa=G(M_\odot+m)$ is the gravitational parameter, with gravity constant $G$.
Using the vis-viva equation $V^2=\kappa (2/r-1/a)$ and assuming an elliptic orbit for the NEO, i.e. $ r= a(1-e^2)/(1+e \cos{\phi})$ we find
\begin{equation}
(\Delta a)^2=\frac{4a^3}{\kappa} \frac{1+2e\cos{\phi}+e^2}{1-e^2}(\Delta V)^2.\label{eggl:eq2}
\end{equation}
It is easily seen from equation (\ref{eggl:eq2}) that a given deflection ($\Delta V$) has the most effect on the semimajor axis when it is applied at 
the NEO's pericenter ($\phi=0$). This is not necessarily always the case, as other trade-offs can be involved \citep{belton-et-al-2011}.
In order to assess the chances of validating a given deflection, we can compare the effective change in semimajor axis with its current 
uncertainty ($\sigma_a$) and define a so-called 'deflection signal-to-noise ratio' (DSNR):
\begin{equation}
DSNR=\sqrt{\frac{(\Delta a)^2}{(\sigma_a)^2}} \label{eggl:eq3}
\end{equation}
Inserting equation (\ref{eggl:eq3}) into (\ref{eggl:eq2}) we arrive at 
\begin{equation}
(\sigma_a)^2=\frac{4a^3}{\kappa} \frac{1+2e\cos{\phi}+e^2}{1-e^2}\frac{(\Delta V)^2}{DSNR^2} \label{eggl:eq4}
\end{equation}
Equation (\ref{eggl:eq4}) yields the necessary constraints on the NEO's semimajor axis to  allow for a detection of a
deflection $\Delta V$ with a signal-to-noise ratio $DSNR$.
In order to be able to detect the previously assumed $\Delta V=1\;cm/s$ for an asteroid with $a=1.05\; au$, $e=0$ at a signal-to-noise ratio of $DSNR=10$,
its orbital uncertainty in semimajor axis should be no larger than $7\cdot10^{-8}\;au$. This translates into permissible observational errors $\ll1''$.
Hence, either high precision astrometry is available before and after the mitigation process, or a radio science experiment much like NEAR is required to guarantee
deflection validation \citep{yeomans-et-al-2000}.

\section{Binary Asteroids}
\begin{figure}[ht!]
 \centering
 \includegraphics[width=0.48\textwidth,clip]{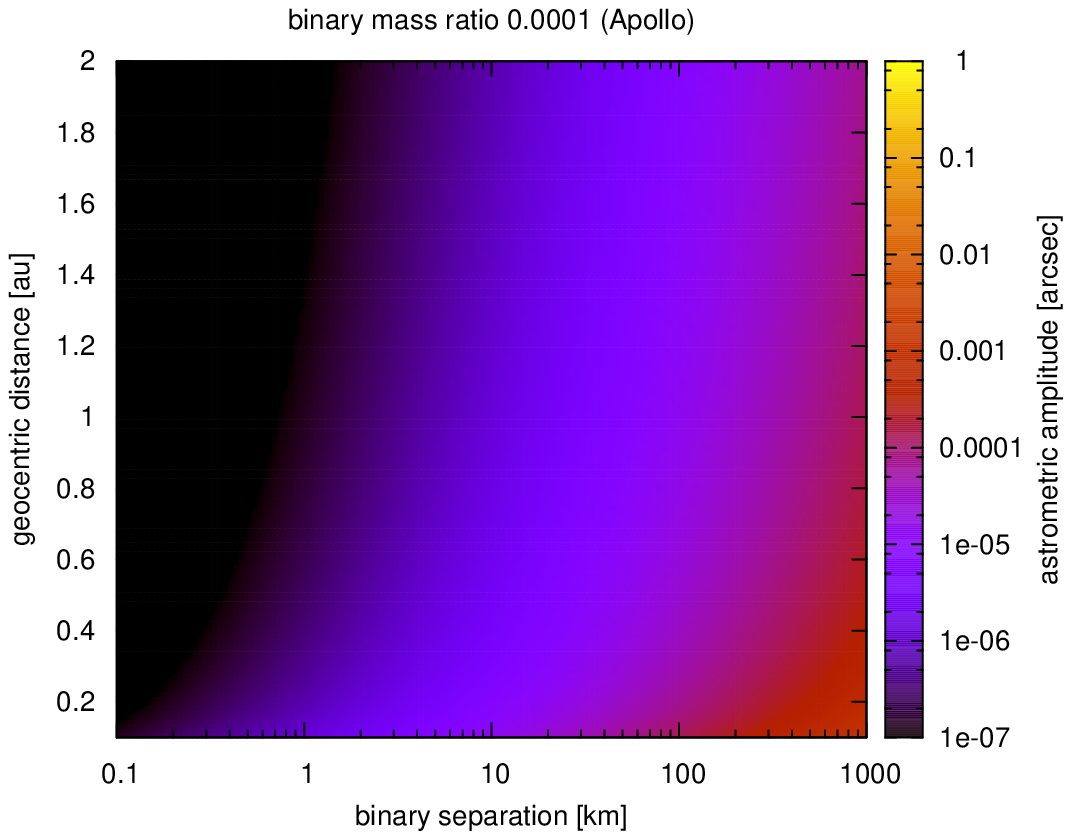}%
 \includegraphics[width=0.48\textwidth,clip]{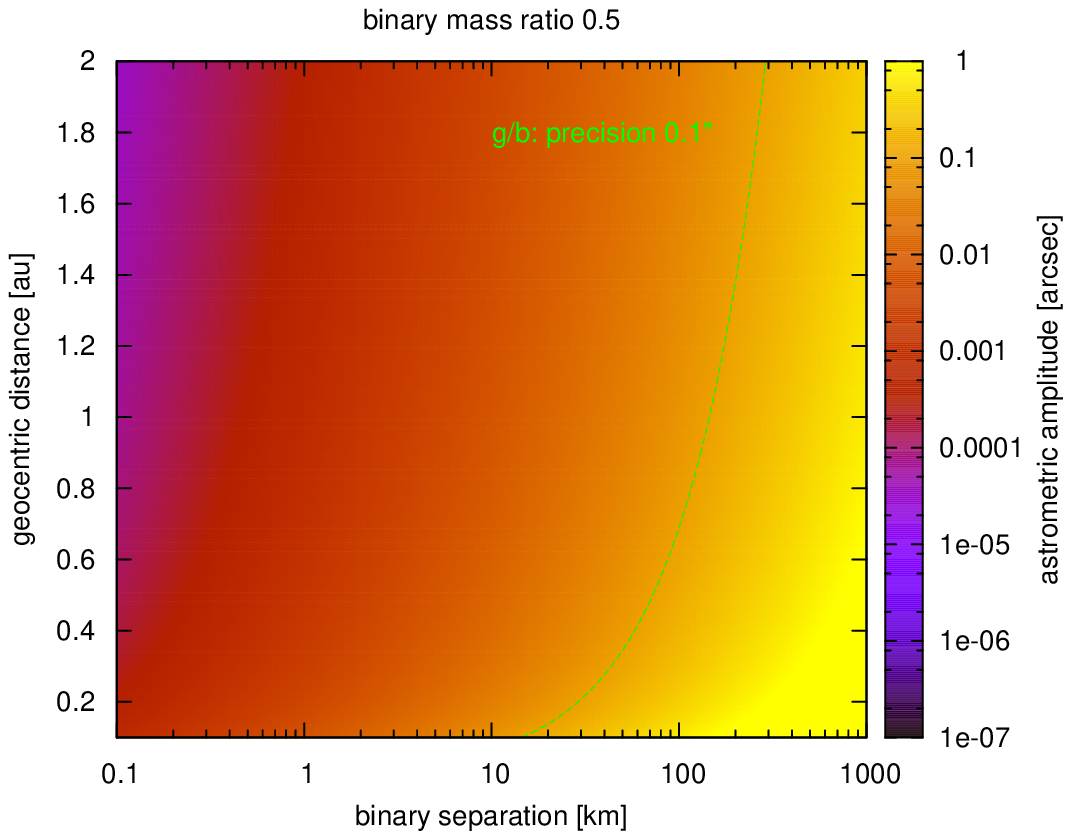}      
  \caption{{\bf Left:} Expected astrometric amplitudes of the center of brightness of the larger asteroid in a binary configuration where one mass is dominant.
  The mass ratio is comparable to that of 1862 Apollo. Contemporary ground-based astrometry is not potent enough to resolve the motion of the center of brightness for such mass ratios.
  {\bf Right:} If both asteroids have comparable masses, however, high precision ground-based astrometry would suffice to cover binary separations $>10-100km$ 
  depending on the asteroid's geocentric distance.}
  \label{eggl:fig34}
\end{figure}
\begin{figure}[ht!]
 \centering
 \includegraphics[width=0.48\textwidth,clip]{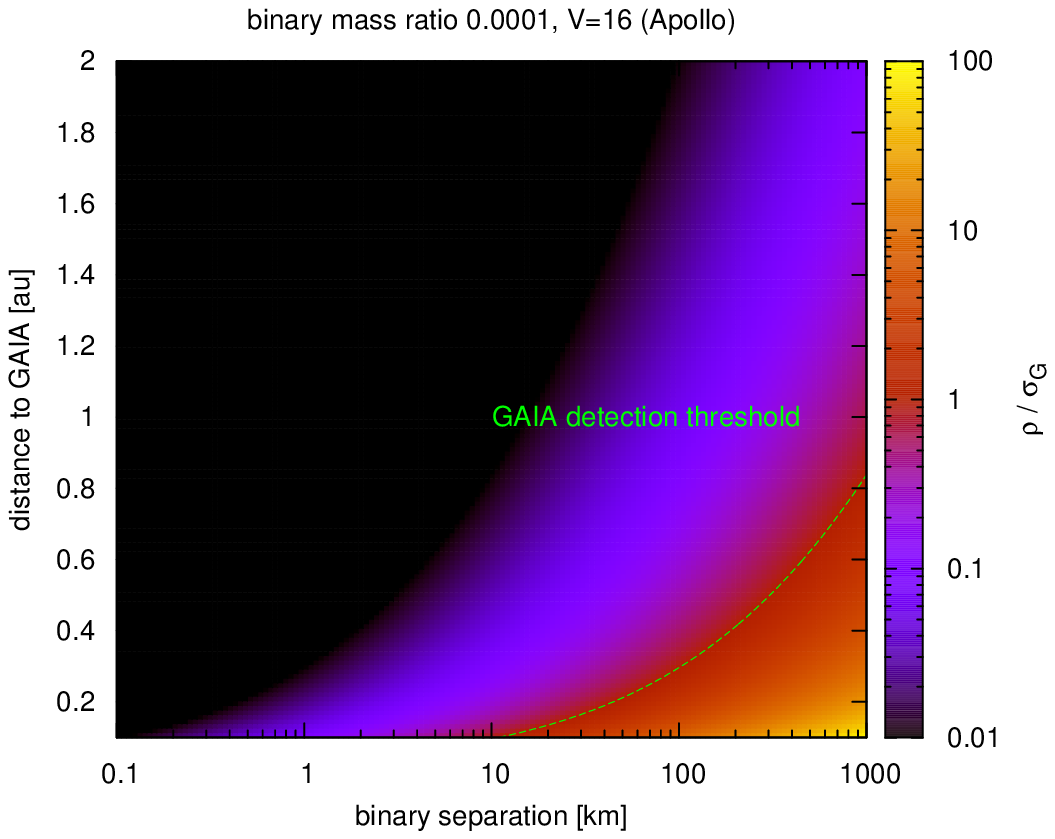}%
 \includegraphics[width=0.48\textwidth,clip]{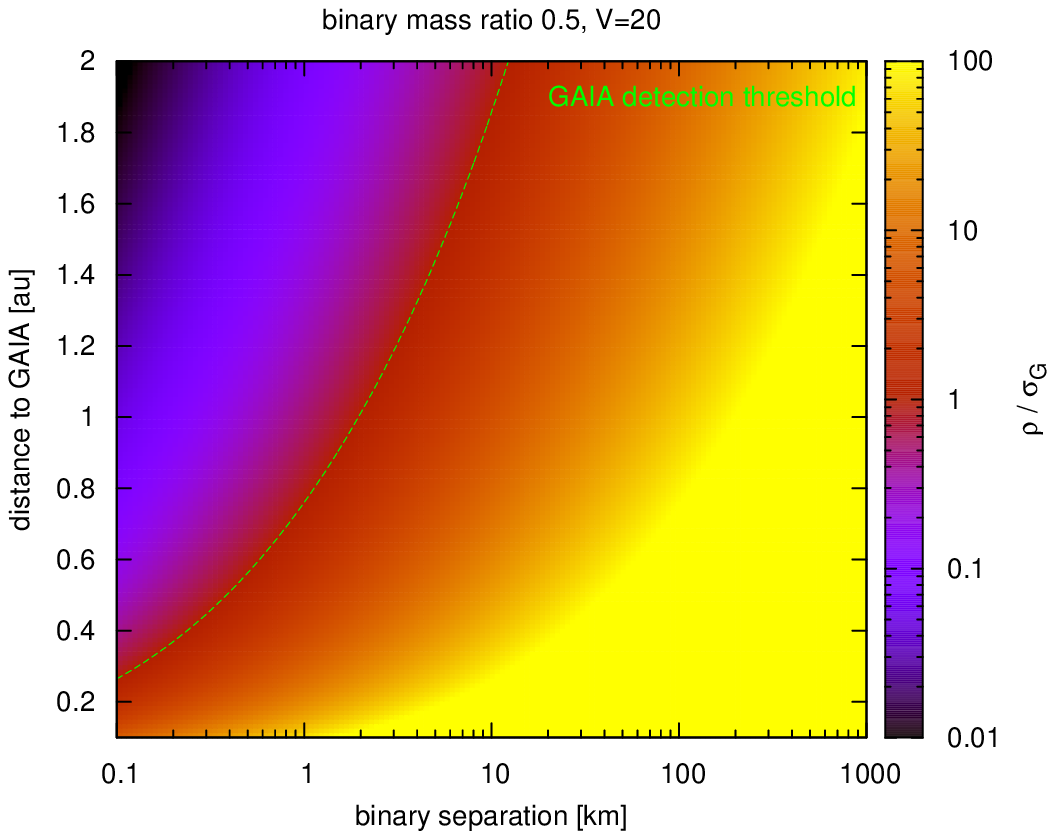}      
  \caption{ Following  \citet{pravec-2012} the ratio of the astrometric amplitude of the brighter asteroid divided by Gaia's astrometric precision is shown as 
  a function of binary separation and the geocentric distance of the binary. Here, we assume an angle of $\pi/2$ between the heliocentric position vectors of the asteroid and Gaia.
  {\bf Left:} A configuration similar to Fig.\ref{eggl:fig34} is shown for the case of Gaia observations. Even Gaia will be hard pressed to detect the astrometric 'wobble'
  of the main component, if the companion is very light. The mass ratio and visual magnitude have been adopted from 1862 Apollo. {\bf Right:} Equally massive binary components have a much better chance of being astrometrically 
  detectable, if they can be optically separated. Even for faint objects with V=20, most of the relevant binary separations are accessible to Gaia.}
  \label{eggl:fig56}
\end{figure}

Discovering shortly before a deflection event that a possible target of a mitigation campaign consists in reality of two separate bodies   
can be considered a rather nasty surprise. Yet, binary asteroids might be favorable targets for
a mitigation demonstration mission \citep[e.g. AIDA/DART][]{cheng-et-al-2012}.
Hence, determining whether a mission target belongs to the roughly 15\% of NEOs that are binary should be a priority \citep{pravec-et-al-2006}.
The variations in the position of the brighter asteroid's photo-center due to its motion around the common center of gravity can be used to 
identify multiplicity and generate binary orbital elements \citep{pourbaix-2002}. 
In order to assess which binary NEO configurations are detectable with standard and high precision astrometry, we estimate the maximum astrometric amplitude ($\rho$) generated
by the mutual orbital motion of binary asteroids following \citet{eggl-et-al-2013}. Assuming a circular mutual orbit we have
\begin{equation}
\rho = \arctan(\mu a/r),
\end{equation} where $a$ is the binary's separation and $r$ the distance to the observer.
Maximum astrometric amplitudes for NEOs with different separations and mass ratios are presented in Fig.~\ref{eggl:fig34}.
If both asteroids have similar masses, i.e. a mass ratio ($\mu$) near 0.5, 
ground-based high precision astrometry can access a range of binary separations  $>10-100km$ depending on the geocentric distance of the NEO. 
Improving ground-based detection of astrometric binaries requires not only better hardware. As mentioned in section \ref{eggl:sec2}, a unified astrometric catalog such as provided
by Gaia will allow for more precise reduction of astrometric data.
In any case, Gaia itself should be able to cover almost all relevant NEO binary separations for equally massive binaries, even if they are faint.
This is shown in Fig.~\ref{eggl:fig56}. Here, we follow \citet{pravec-2012} and consider the astrometric amplitude of the wobble of the binary's components 
divided by the expected precision of Gaia, with a detection threshold of $\rho/\sigma_G>1$. Gaia's expected precision is a function of the apparent brightness (V), i.e.
\begin{equation}
\sigma_G=0.15\cdot10^{0.2(V-16)},
\end{equation}
where $\sigma_G$ [mas] is the standard uncertainty of a single-epoch measurement of Gaia \citep{pravec-2012}.
Furthermore, a constant angle of $\pi/2$ between the object and Gaia is assumed. 
For asteroids with very small companions, such as 1862 Apollo, however, one can see in the left panel of Fig.~\ref{eggl:fig56} that even Gaia might be hard pressed to detect the astrometric oscillations of the center of brightness 
  of the main component, especially if the binary is close.
  Also, it can be rather difficult to disentangle astrometric signals from other photo-center offsets due to irregular shapes and rotation of the observed object \citep[e.g.][]{ortiz-et-al-2011}, and besides to disentangle from other photo-center offsets due to rotation, shape and phase.
  Hence, 'asteroid moons' might remain undetected in a preliminary threat assessment study, especially if the warning time is short. 
  This fact should be considered in the design of a mitigation mission. A small explorer S/C sent prior to an impactor, for instance, 
  could help to avoid an accidental collision of the impact vehicle with the smaller satellite \citep{cheng-et-al-2012}.

\section{Conclusions}
Acquiring high precision astrometry of NEOs is essential for asteroid threat
assessment and mitigation. High quality data can be used to reduce
orbit uncertainties, which will allow for quick and precise estimates of impact
probabilities.
We could show that high precision astrometry is necessary, if 
the success of an asteroid deflection demonstration mission is to be evaluated via ground-based observations.
Furthermore, knowing whether an asteroid is multiple is an essential part of a NEO threat assessment, as it permits an adaption of mitigation strategies and, thus
safeguards us against the consequences of deflection failures.
A determination of asteroid multiplicity via high precision
astrometry seems also feasible with Gaia, at least for binaries with components of similar mass. 
Even current ground-based facilities might be able to contribute, if the binary has separations larger than 10km and multiple close encounters with the Earth.

Based on observational residuals of NEOs published by the IAU MPC we estimated the current average astrometric performance of ground-based observations 
to lie between $0.13''$ and $2''$. High precision measurements constituted less than 2\% of the total number of observations in 2013. 
Even considering pessimistic scenarios, ESA's Gaia mission should
produce astrometric results for asteroids with a visual magnitude $V<20$ that are orders of magnitude beyond the current average. 
However, Gaia's mission lifetime is restricted to 5 years. While the astrometric catalog provided by the Gaia mission 
will allow for a permanent improvement of ground-based observation accuracy, an astrometric precision comparable to the in-mission performance will be hard to sustain.
A long term follow up strategy should be developed that can continue Gaia's astrometric legacy.

\begin{acknowledgements}
The authors would like to acknowledge the support of the NEOShield project
funded by the European Union Seventh Framework Program (FP7/2007-2013)
under grant agreement no. 282703. A. Ivantsov would like to appreciate the financial support 
at IMCCE-Observatoire de Paris through the program 'Research in Paris 2012'.
D. Bancelin expresses his gratitude towards Austrian FWF projects AS11608-N16 and P22603-N16. 
This research has made use of data provided by the IAU Minor Planet Center.
\end{acknowledgements}

\bibliographystyle{aa}  
\bibliography{eggl_neo} 

\end{document}